\documentclass[aps,showpacs,twocolumn,twoside]{revtex4}
\usepackage{epsfig}
\begin{document}

\def\ket#1{|#1\rangle}
\def\bra#1{\langle#1|}
\def\scal#1#2{\langle#1|#2\rangle}
\def\matr#1#2#3{\langle#1|#2|#3\rangle}
\def\keti#1{|#1)}
\def\brai#1{(#1|}
\def\scali#1#2{(#1|#2)}
\def\matri#1#2#3{(#1|#2|#3)}
\def\sloup#1#2{\left(\begin{array}{c}#1\\#2\end{array}\right)}

\title{Evolution of spectral properties along the O(6)-U(5) transition \\ 
       in the interacting boson model. I. Level dynamics}
\author{
Stefan Heinze$^1$, Pavel Cejnar$^2$, Jan Jolie$^1$,  Michal Macek$^2$}

\affiliation{
$^1$Institute of Nuclear Physics, University of Cologne,
Z\"ulpicherstrasse 77, 50937 Cologne, Germany\\
$^2$Faculty of Mathematics and Physics, Charles University,
V Hole{\v s}ovi{\v c}k{\'a}ch 2, 180\,00 Prague, Czech Republic
}

\date{\today}
\begin{abstract}
We investigate the evolution of quantal spectra and the corresponding
wave functions along the [O(6)$-$U(5)]$\supset$O(5) transition of the 
interacting boson model.
The model is integrable in this regime and its ground state passes 
through a second-order structural phase transition.
We show that the whole spectrum as a function of the Hamiltonian 
control parameter, as well as structures of all excited states, 
exhibit rather organized and correlated behaviors, that provide 
deeper insight into the nature of this transitional path.
\pacs{21.60.Fw, 21.10.Re, 05.45.Mt}
\end{abstract}

\maketitle

\section{Introduction}

Properties of the interacting boson model (IBM) \cite{Iach} in transitional 
regimes between various dynamical symmetries have been extensively studied 
mainly in connection with zero-temperature quantum phase transitions 
\cite{Gil,Feng}. 
In any of such transitions, the structures of the ground state and few 
low-lying states change abruptly (for the system size tending to infinity)
at a certain critical point, located on the way between the two
dynamical-symmetry limits, see, e.g., 
Refs.~\cite{Diep,Mor,Jol1,Jol2,Cej4,Cej5}.
This behavior finds experimental evidence (in a finite-$N$ approximation) 
in observed variations of nuclear shapes in some isotopic or isotonic 
chains of nuclei.

The IBM phase transitions are of the first order, except the isolated point
of a second-order transition, which is located at the intersection of 
borders between spherical and deformed, and between prolate and oblate 
shapes in the parameter space \cite{Diep,Mor,Jol1}.
To pass this point, one commonly starts from the O(6) dynamical 
symmetry and proceeds to U(5) via the line of unbroken O(5) dynamical 
symmetry.
The deformed-to-spherical second-order phase transition on this path
manifests itself as a nonanalytic but continuous change of the 
ground-state deformation, in contrast to the discontinuous changes 
observed along the other (even infinitely close) transitional paths.
This type of phase structure of the parameter space agrees with the 
classical Landau theory of thermodynamic phase transitions, that is 
applicable at zero temperature if the role of thermodynamic variables 
is taken by the model control parameters \cite{Jol2,Cej4}, and
with catastrophe theory \cite{Gil,Mor}. 

The above-mentioned [O(6)$-$U(5)]$\supset$O(5) transitional path differs 
from the others also in that it does not destroy the integrability of the 
Hamiltonian.
Due to the underlying O(5) dynamical symmetry \cite{Lev}, the integrals
of motions along the whole transition form a complete set of commuting 
operators and the Hamiltonian eigenproblem can be solved analytically 
\cite{Pan,Duk}.
This was used for an explicit calculation of some second-order 
phase-transitional observables \cite{Ari}.
Recent studies of the O(6)-U(5) transitional path were also based on 
the concepts of the E(5) critical-point dynamical symmetry \cite{Ia2,Ari2} 
and the quasidynamical symmetry \cite{Row}.
 
The ultimate mechanism that is on the deepest level responsible for 
the occurrence of ground-state phase transitions of various orders in 
quantum many-body systems remains unclear.
The distinction between the IBM first- and second-order phase 
transitions, e.g., was shown \cite{Cej5} to be connected with different 
densities of unavoided energy crossings (branch points) in the 
complex-extended parameter space, which in the $N\to\infty$ limit 
accumulate infinitely close to critical points on the real axes 
(in analogy with similar behaviors of complex zeros of partition 
functions in thermodynamic phase-transitional systems).
However, many questions---among them the role of integrability in the
process of dynamical-symmetry breaking---still remain open.

The present work contributes to the mapping of this relatively new 
territory of physics by studying in detail various spectral observables 
associated with the integrable phase-transitional path in the IBM.
In particular, we investigate the evolution of energies and wave 
functions of individual Hamiltonian eigenstates with zero angular 
momentum along the whole [O(6)$-$U(5)]$\supset$O(5) line. 
It is shown that this transitional class of IBM exhibits rather 
peculiar features.

We will combine two totally different, but mutually related general 
approaches: 
(i) the theory of level dynamics, initiated by Pechukas and Yukawa 
\cite{Pec}, also known as the dynamical Coulomb-gas analogy, and 
(ii) the semiclassical theory of quantal spectra, represented by the 
Gutzwiller and Berry-Tabor trace formulas \cite{Gut}.
Results obtained by applying both these approaches will be presented
in two parts: approach (i) is discussed in the present article 
(Part~I), which gives numerical results on level dynamics, while 
approach (ii) will be used in the following article \cite{partII} 
(Part~II).

The Pechukas-Yukawa theory describes the dynamics of individual 
levels and the interaction matrix elements via a set of coupled 
differential equations, where the varying Hamiltonian control 
parameter plays the role of time.
It enables one to understand the evolution of spectral observables
with the control parameter (including eventual phase transitions) 
in a more intuitive way, using the parallel with a classical ensemble 
of charged particles moving in one dimension.

The trace formulas, on the other hand, describe a snapshot of the 
energy spectrum at each fixed value of the control parameter
(\lq\lq time\rq\rq) by expressing the quantum density of states (as 
a function of energy) through properties of periodic orbits in the 
classical limit of the system.
Since both methods (i) and (ii) translate the original problem to the 
classical language, they often provide deeper understanding of 
specific behaviors observed on the quantum level.
Also in our case, the most significant features of the IBM in the 
[O(6)$-$U(5)]$\supset$O(5) transitional regime will be elucidated 
by both kinds of classical concepts involved in the above approaches.

The plan for this part of the paper is the following: 
In Section \ref{hami} we will briefly describe the quantum Hamiltonian
under study, its integrals of motions and phase-transitional features.
Section \ref{dyn} presents numerical results on the level dynamics 
and their interpretation in the framework of the Pechukas-Yukawa 
theory.
The accompanying changes in the structure of wave functions are then
discussed in Section \ref{state}.
Finally, Section \ref{conclu} contains partial conclusions of this
part of the paper.

\section{Quantum Hamiltonian}
\label{hami}

The interacting boson model \cite{Iach} describes shapes and collective 
motions of atomic nuclei in terms of an ensemble of $N$ interacting $s$ 
and $d$ bosons with angular momenta 0 and 2, respectively.
To analyze the evolution of properties of this model along the 
O(6)-U(5) transitional path, we adopt the 
Hamiltonian
\begin{equation}
\hat{H}(\eta)=a\left[-\frac{1-\eta}{N^2}(\hat{Q}\cdot\hat{Q})+
\frac{\eta}{N}\,\hat{n}_{d}\right]\ ,
\label{ham}
\end{equation}
where the dimensionless control parameter $\eta\in[0,1]$ changes 
the proportion of both competing terms and drives the system 
between the O(6) ($\eta=0$) and U(5) ($\eta=1$) dynamical symmetries. 
The operator $\hat{n}_{d}=(d^{\dagger}\cdot{\tilde d})=N-\hat{n}_s$ 
represents the $d$-boson number, while $\hat{Q}\equiv\hat{Q}_0^{(2)}=
[s^{\dagger}{\tilde d}+d^{\dagger}{\tilde s}]^{(2)}$ stands for the 
O(6)-U(5) quadrupole operator.
The energy scale is set by an arbitrary factor $a$, which in the 
following will be fixed at the value $a=1$~MeV. 

Hamiltonian (\ref{ham}) is a special case of a more general Hamiltonian 
of the same form, but with the quadrupole operator given by 
$\hat{Q}_\chi^{(2)}=\hat{Q}_0^{(2)}+\chi[d^{\dagger}{\tilde d}]^{(2)}$, 
where $\chi\in[-\frac{\sqrt{7}}{2},+\frac{\sqrt{7}}{2}]$ is an additional 
control parameter.
Eq.~(\ref{ham}), where $\chi=0$, can be decomposed \cite{Cej6} into 
a linear combination of Casimir invariants corresponding to the O(6), 
O(5), O(3), and U(5) algebras, with no admixture of SU(3), 
$\overline{\rm SU(3)}$, and $\overline{\rm O(6)}$ invariants, i.e., 
describes the [O(6)$-$U(5)]$\supset$O(5) transitional line in the 
extended Casten triangle \cite{Jol1}.

The above Hamiltonian can also be rewritten as
\begin{equation}
\hat{H}(\eta)=(1-\eta)\hat{H}(0)+\eta\hat{H}(1)=\hat{H}_0+\eta\hat{V}\ , 
\label{hamgen}
\end{equation}
which is the form well known from various studies of quantum phase 
transitions.
Assuming $a=1$, we obtain 
\begin{eqnarray}
\hat{H}_0 & = & -\frac{1}{N^2}(\hat{Q}\cdot\hat{Q})\ ,
\label{H0}
\\
\hat{V}   & = & \frac{1}{N}\,\hat{n}_{d}+\frac{1}{N^2}(\hat{Q}
\cdot\hat{Q})\ .
\label{V}
\end{eqnarray}
The evolution of Hamiltonian (\ref{hamgen}) with $\eta$ can be 
treated in a perturbative way since $\hat{H}(\eta+\delta\eta)=
\hat{H}(\eta)+{\delta\eta}\,\hat{V}$.
Note that the powers of $N$ in denominators of Eqs.~(\ref{ham}), 
(\ref{H0}) and (\ref{V}) guarantee convenient scaling of the Hamiltonian 
with variable boson number $N\gg 1$.

It can be easily shown that for Hamiltonian (\ref{hamgen}) the 
ground-state average $\langle V\rangle_{\eta}\equiv\matr{\psi_1(\eta)}
{\hat{V}}{\psi_1(\eta)}=\frac{dE_1(\eta)}{d\eta}$ [where $E_1(\eta)$ and
$\ket{\psi_1(\eta)}$ are the ground-state energy and wave function, 
respectively] is a nonincreasing function of $\eta$.
Therefore, if $\hat{V}$ is nonnegative---as in our specific case, see 
Eq.~(\ref{V})---then an instantenous satisfaction of $\langle V\rangle
_{\eta_{\rm c}}=0$ at some critical point $\eta_{\rm c}$ implies that 
the average gets fixed for all $\eta\geq\eta_{\rm c}$, freezing both 
the energy and wave function of the ground state.
At this point, the system may exhibit (for $N\to\infty$) a ground-state 
phase transition of order $\kappa\geq 2$.
If the second derivative of energy changes discontinuously from 
a value $\frac{d^2E_1(\eta)}{d\eta^2}=\frac{d\langle V
\rangle_{\eta}}{d\eta}<0$ at $\eta=\eta_{\rm c-}$ to zero at 
$\eta=\eta_{\rm c+}$, the transition is of the second order.
Higher-order transitions \cite{Gil,Feng} would require additional 
constraints, namely $\frac{d^k\langle V\rangle_{\eta}}{d\eta^k}
\bigr|_{\eta_{\rm c}-}=0$ for $k<\kappa$.

It is not difficult to see that for the specific Hamiltonian in 
Eq.~(\ref{ham}) the ground-state average of $\hat{V}$ indeed 
interpolates between a positive value at $\eta=0$ and zero at $\eta=1$.
However, the phase-transitional scenario is generically allowed only 
in the limit of infinite Hilbert-space dimensions, thus $N\to\infty$, 
when the ground-state energy as a function of $\eta$ may acquire 
nonanalytic character.
The asymptotic critical point is located at $\eta_{\rm c}=\frac{4}{5}
=0.8$.
At this point, the deformed ground-state 
configuration, given by a mixed-boson condensate $\ket{\psi_1}\propto
(s^{\dagger}+\beta_{\rm gs}d_0^{\dagger})^N\ket{0}$, changes into the 
pure $s$-boson condensate, $\ket{\psi_1}\propto(s^{\dagger})^N\ket{0}$, 
characterizing the spherical U(5) phase.
In the left vicinity of the critical point the ground-state
\lq\lq deformation parameter\rq\rq\ $\beta_{\rm gs}$ drops to zero as 
$\beta_{\rm gs}\propto\sqrt{\eta_{\rm c}-\eta}$ \cite{Jol2} and the 
corresponding value of $\langle V\rangle_{\eta}$ behaves according 
to $N\to\infty$ asymptotic formula 
$\langle V\rangle_{\eta}\propto(\eta_{\rm c}-\eta)$ \cite{Cej5}.
Thus both $\beta_{\rm gs}$ and $\langle V\rangle_{\eta}$ can be considered 
as order parameters describing a second-order quantum phase transition, 
$\kappa=2$, with critical exponents $\frac{1}{2}$ and 1, respectively.

The limits $\hat{H}(0)$ and $\hat{H}(1)$ of Eq.~(\ref{ham}) posses the
O(6) and U(5) dynamical symmetries, respectively.
Since the dynamical-symmetry Hamiltonians are constructed using
solely observables \lq\lq in involution\rq\rq\ (the Casimir 
invariants of the respective algebraic chain), they are always 
integrable \cite{Zhang}.
Moreover, because the O(5) dynamical symmetry underlying both
O(6) and U(5) limits is not broken in the transitional regime, the 
integrability of Hamitonian (\ref{ham}) is preserved for all 
values of $\eta$ \cite{Alha2,Cej6}.
Indeed, one can find five mutually commuting integrals of motion,
the same number as the dimension of the classical configuration 
space (given by two geometric parameters and three Euler angles
\cite{Hatch}).
Four of these integrals can be associated with the following
quantum numbers: energy $E_i(\eta)$ given by the Hamiltonian
$\hat{H}(\eta)$, squared angular 
momentum $l(l+1)$ represented by $\hat{L}^2$ (where $\hat{L}=
[d^{\dagger}{\tilde d}]^{(1)}$), its projection $m$ determined 
from $\hat{L}_z$, and the seniority $v$ defined through the $v(v+3)$ 
eigenvalue of the O(5) Casimir invariant \cite{Iach}:
\begin{equation}
\hat{C}_2[O(5)]=\frac{1}{5}(\hat{L}\cdot\hat{L})+
2(\hat{T}_3\cdot\hat{T}_3)
\label{senior}
\end{equation}
(where $\hat{T}_3=[d^{\dagger}{\tilde d}]^{(3)}$). 
The fifth integral of motion, connected with the so-called missing
label ${\tilde n}_{\Delta}$ of the O(5)$\supset$O(3) reduction, 
is not given explicitly, but its existence is guaranteed by the fact 
that there must be five independent commuting operators in the 
complete set, so the Hamiltonian (which is made of four of them) 
commutes with the fifth one \cite{Zhang}.
Note that in this paper we will only consider the set of states 
with zero angular momentum, $l=0$.

\section{Eigenvalue dynamics}
\label{dyn}

\subsection{Pechukas-Yukawa equations}
\label{pec}

Drawing the dependence of all individual level energies $E_i(\eta)$ 
for Hamiltonian (\ref{hamgen}) on the control parameter, 
one obtains a picture containing $n$ continuous curves that resemble
trajectories $x_i(t)$ of an ensemble of particles in one dimension.
The motion of levels is described by a set of Hamilton-type first-order 
differential equations associated with a gas of particles interacting 
via two-dimensional Coulomb force:
\begin{equation}
\frac{d^2E_i}{d\eta^2}=2\sum_{j(\neq i)}
\frac{|V_{ij}|^2}{E_i-E_j}
\label{force}
\end{equation}
(analogous to $\frac{d^2x_i}{dt^2}=\frac{1}{2\pi\epsilon_0}
\sum_{j(\neq i)}\frac{q_iq_j}{x_i-x_j}$).
In contrast to the ordinary gas dynamics, however, the \lq\lq product 
charge\rq\rq\ $|V_{ij}|^2=|\matr{\psi_i(\eta)}{\hat{V}}{\psi_j(\eta)}|^2
\leftrightarrow q_iq_j\equiv Q_{ij}$ cannot be factorized and varies as the 
\lq\lq time\rq\rq\ $\eta\leftrightarrow t$ elapses. 
Thus the product charges (alias interaction matrix elements) are also 
dynamical variables, subject to specific evolution, and the system's 
phase space is larger than $2n$.
Besides Eq.~(\ref{force}) we have
\begin{equation}
\frac{dV_{ij}}{d\eta}=-\frac{V_{ii}-V_{jj}}{E_i-E_j}+\sum_{k(\neq i,j)}
\frac{E_i+E_j-2E_k}{(E_i-E_k)(E_j-E_k)}V_{ik}V_{kj}
\ .
\label{charge}
\end{equation}
for $i\neq j$, and
\begin{equation}
\frac{dE_i}{d\eta}=V_{ii}\ .
\label{veloc}
\end{equation}

Eqs.~(\ref{force})--(\ref{veloc}) are equivalent to the well-known
Pechukas-Yukawa set of equations \cite{Pec}, although we use here 
a slightly different form than the one usually found in textbooks 
\cite{Stock}.
The system described by these equation is deterministic and even 
integrable.
If all energies and interaction matrix elements are known at a single 
point $\eta$ (for instance $\eta=0$), the equations determine $E_i$'s 
and $V_{ij}$'s for all other $\eta$ values.

Since the product charge $|V_{ij}|^2$ in Eq.~(\ref{force}) is
nonnegative, the levels never touch each other unless their 
mutual interaction completely vanishes.
In absence of symmetry-dictated zeros of the interaction matrix,
the coincidence of simultaneous convergences $|V_{ij}|^2\to 0$ and 
$E_{i+1}-E_i\to 0$ is extremely unlikely, which gives rise to the 
well-known \lq\lq no-crossing\rq\rq\ rule for level energies.
The presence of symmetries, however, induces the disappearance of 
$V_{ij}$'s for certain sets of states which, therefore, can cross.
In case of Hamiltonian (\ref{ham}), this concerns levels with
different values of angular momentum $l$ and levels with different 
seniority $v$.

\subsection{Level bunching around $E\approx 0$}
\label{bunch}

\begin{figure}
\epsfig{file=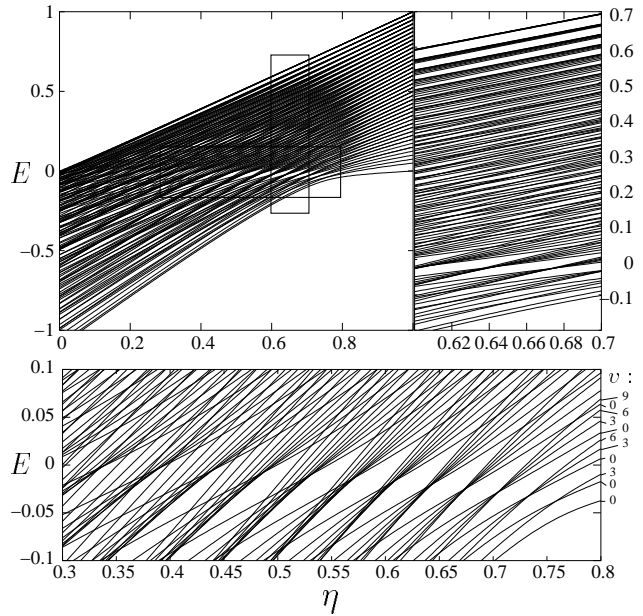,width=\linewidth}
\caption{\protect\small Spectrum of Hamiltonian (1) with $N=40$ as
 a function of $\eta$ for $l=0$ levels with all seniorities. The
 vertical and horizontal rectangles are expanded in the upper right
 and lower panels, respectively. The seniority is assigned to several
 levels in the lower panel.}
\label{levall}
\end{figure}

In Figure~\ref{levall} we show the dynamics of all levels with $l=0$
along the $\eta\in[0,1]$ path between the O(6) and U(5) dynamical 
symmetries of Hamiltonian (\ref{ham}). 
The calculation was performed by numerical diagonalization of the
Hamiltonian for $N=40$ bosons. 
One can observe numerous level crossings, particularly in the region 
around $E\approx 0$ (the horizontal rectangle, expanded in the lower 
panel), which is a consequence 
of the unbroken O(5) dynamical symmetry of the system.
Indeed, the seniority quantum numbers, as marked for few levels on
the rightmost side of the lower panel, differ for any pair of levels 
that cross at some point.
We will see below that the crossings disappear after separation 
of levels with different seniorities into several figures.

The pattern of consecutive compressions and dilutions of the spectrum 
in the region around $E\approx 0$ is one of the most apparent 
attributes of Fig.~\ref{levall} (see the lower panel).
A striking feature of this pattern is the regular sequence that
characterizes the total numbers of levels involved in individual
bunches: when descending from $\eta=0.8$ to $\approx 0.4$, the
sequence goes like 1, 2, 3, 4, \dots
At first, the different seniority states seem to cross exactly at the 
same point (within the available numerical precision), but with
$\eta$ descending below 0.65 the higher seniorities get increasingly 
out of focus, and the bunching pattern becomes more and more 
diffuse.
Nevertheless, the structure of alternating clusters and gaps 
extends over a wide range $\eta\in[0.3,0.8]$.
Secondary \lq\lq interference\rq\rq\ patterns are also visible at 
other energies (see the vertical rectangle of Fig.~\ref{levall}, 
extended in the upper right-hand-side panel), but these are much 
weaker than the main one.

The energy $E\approx 0$, where the bunching pattern appears, is 
significant because it corresponds to the local maximum at 
$\beta=0$ of the classical potential \cite{Iach,Hatch} corresponding 
to Hamiltonian (\ref{ham}).
The bunching of levels thus develops just at the value of energy
where the classically accessible range of the deformation parameters,
$\beta\in[\beta_{\rm min},\beta_{\rm max}]$, extends due to
$\beta_{\rm min}$ becoming zero.
The connection of the bunching pattern with the IBM classical 
dynamics will be elaborated in Part II of this contribution 
\cite{partII}.

\begin{figure}
\epsfig{file=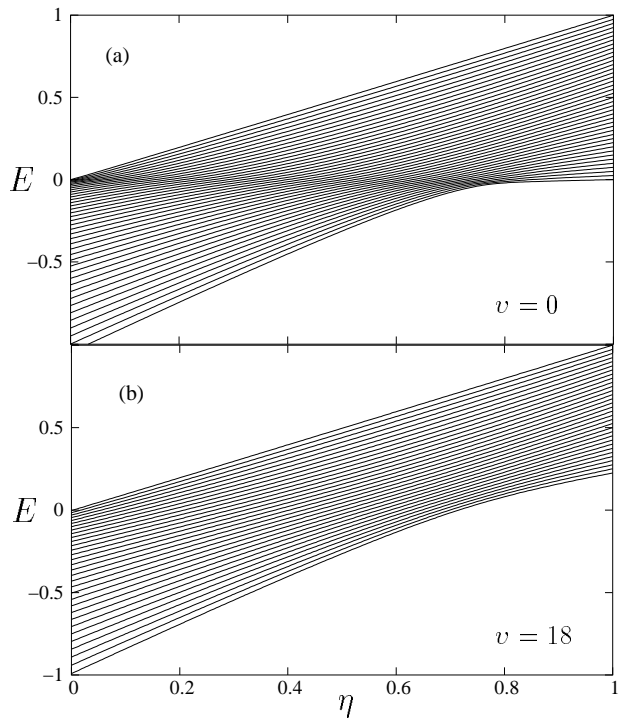,width=\linewidth}
\caption{\protect\small Spectrum of Hamiltonian (1) with $N=80$ for 
 $l=0$ levels with seniority $v=0$ (panel a) and $v=18$ (panel b).} 
\label{levsen}
\end{figure}

\subsection{Shock-wave scenario}
\label{shock}

\begin{figure}
\epsfig{file=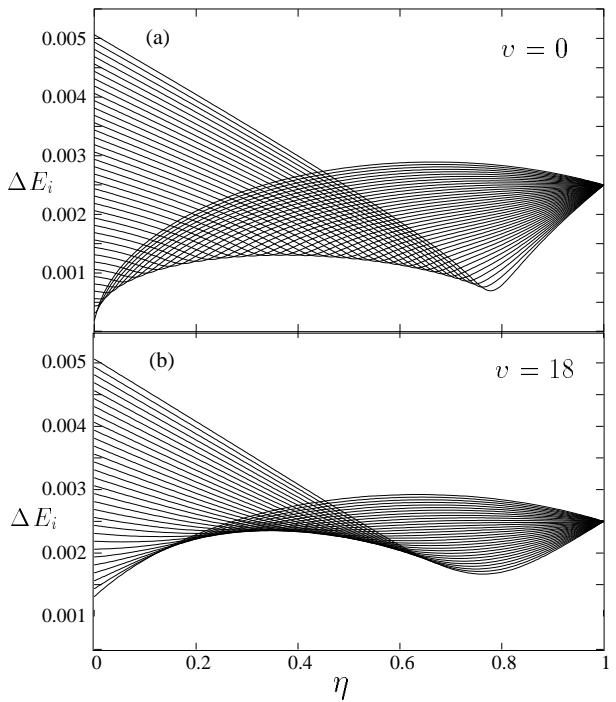,width=\linewidth}
\caption{\protect\small Distances of neighboring levels from 
 Fig.~\ref{levsen}.}
\label{distan}
\end{figure}

Figure~\ref{levsen} demonstrates that the level bunching pattern
can be deconvoluted by separating states with different seniorities.
Here we show the level dynamics for $v=0$ (this set includes the 
ground state) and $v=18$, with the boson number $N=80$.
Clearly, the $v=0$ levels in panel (a) form a smooth flow with 
a \lq\lq shock wave\rq\rq\ propagating from the top of the spectrum 
(at $\eta=0$) to the ground state (at $\eta=0.8$).
The mutual distances $\Delta E_i=E_{i+1}-E_i$ of individual $v=0$ 
levels as functions of $\eta$ are shown in Figure \ref{distan}(a), 
where we can clearly identify points of the closest approach of 
neighboring states as the wave propagates through the ensemble.
Note that due to the energy denominator in Eq.~(\ref{force}), 
a minimal spacing of levels tends to induce maximal 
\lq\lq force\rq\rq\ acting on the relevant levels, which is basically 
the mechanism that keeps the wave moving.
This is also why the wave initiates in the upper (densest) part of
the spectrum (at $\eta=0$, the distance of nearest levels linearly 
decreases with $i$, while at $\eta=1$ it is constant, cf. 
Fig.~\ref{distan}). 

On the other hand, the dynamics of the $v=18$ levels, shown in
panels (b) of both figures, exhibits much weaker interactions.
The flow in Fig.~\ref{levsen}(b) looks almost laminar and the 
minimal distances in Fig.~\ref{distan}(b) (still disclosing
interactions) are about twice larger than in the $v=0$ case.
It can be checked that the weakening of level interactions proceeds 
gradually as $v$ increases.

The shock-wave interpretation of Fig.~\ref{levsen}(a) is particularly 
appealing if used as a tentative reasoning for the ground-state phase 
transition at $\eta_{\rm c}=\frac{4}{5}$.
It seems that this transition results from a highly ordered sequence
of structural changes that propagate from upper to lower parts of the 
spectrum and terminate at the ground state just at the critical point.
This mechanism, however, needs to be verified by an analysis of wave
functions and will be further discussed in Sec.~\ref{state}.

\begin{figure}
\epsfig{file=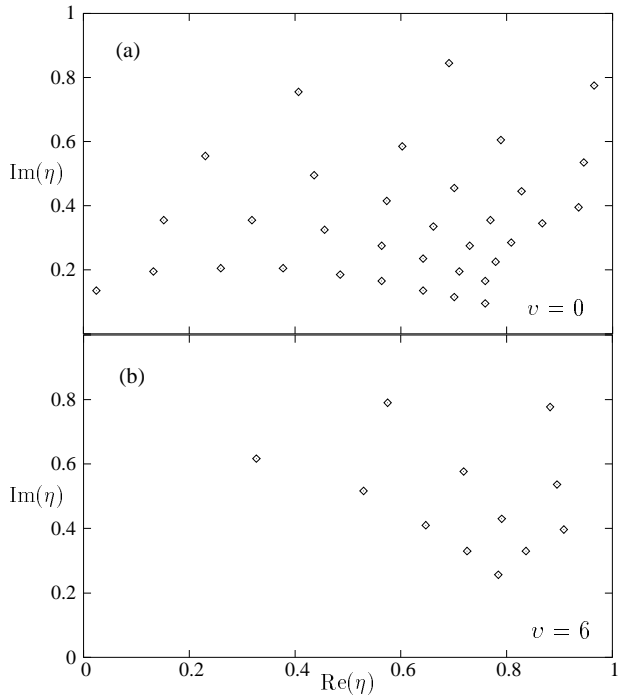,width=\linewidth}
\caption{\protect\small Branch points of Hamiltonian (\ref{ham}) with 
 $N=20$ for $l=0$ states with seniorities $v=0$ (panel a) and $v=6$ 
 (panel b).}
\label{branch}
\end{figure}

Closely related to the regular evolution of level energies is 
the organized pattern of the Hamiltonian branch points in the
complex plane of parameter $\eta$ \cite{Heis}. 
It is shown for $N=20$ in Figure~\ref{branch} for (a) $v=0$ and 
(b) $v=6$.
Branch points are places in the complex-extended parameter 
space where two (or more) Hamiltonian complex eigenvalues become 
degenerate \cite{Kato}.
A branch point located on the real $\eta$-axis would imply a real 
crossing of the corresponding levels, which does not typically happen 
(for levels with different symmetry quantum numbers).
On the other hand, if a given branch point is not on, but sufficiently 
close to the real axis, one observes an avoided crossing of the relevant 
levels at the corresponding value of $\eta$.
Remind that a sequence of such avoided crossings is significant for 
the \lq\lq shock wave\rq\rq.
A cumulation of branch points in infinitesimal
vicinity (for $N\to\infty$) of the critical point $\eta_{\rm c}$ was 
recently shown \cite{Cej5} to constitute the essential triggering 
mechanism for the IBM quantum phase transitions of both orders.

For each seniority, there are altogether $\frac{n(n-1)}{2}$ complex
conjugate pairs of branch points, where $n$ is the dimension of the
given seniority subspace.
Because of numerical constraints, we can only show results for moderate
dimensions that correspond to the lower boson number $N=20$.
As can be seen in Fig.~\ref{branch}, branch points for both
seniorities form rather regular patterns.
In the $v=0$ case (panel a) we notice a chain of points at $\eta<0.8$ 
that approach close to the real axis.
These points clearly correspond to the sequence of avoided crossings 
shown (for a higher boson number) in Fig.~\ref{levsen}(a).
With increasing seniority, the pattern gets more and more separated
from the real axis (see the example in panel b), which results in 
a weakening of level interactions, as observed (for different values
of $N$ and $v$) in Fig.~\ref{levsen}(b).
Note that such an organized behavior of branch points is a privilege 
of only the [O(6)$-$U(5)]$\supset$O(5) transitional class, where the 
separation of seniorities is possible (cf. Ref.~\cite{Cej5}).

\subsection{Focal point and spectral invariant}
\label{focal}

A more detailed view of Fig.~\ref{levsen}(a) discloses that almost all 
$v=0$ levels on the $\eta=0$ side (except perhaps few ones on the top of 
the spectrum) point to a virtually sharp focus on the $\eta=1$ side. 
Indeed, an unperturbed evolution (with no mutual interactions between
levels) would lead to crossing of individual lines at the point 
$(\eta,E)=(1,\frac{1}{2})$, which we call an (approximate) focal point 
of the [O(6)$-$U(5)]$\supset$O(5) transition.

\begin{figure}
\epsfig{file=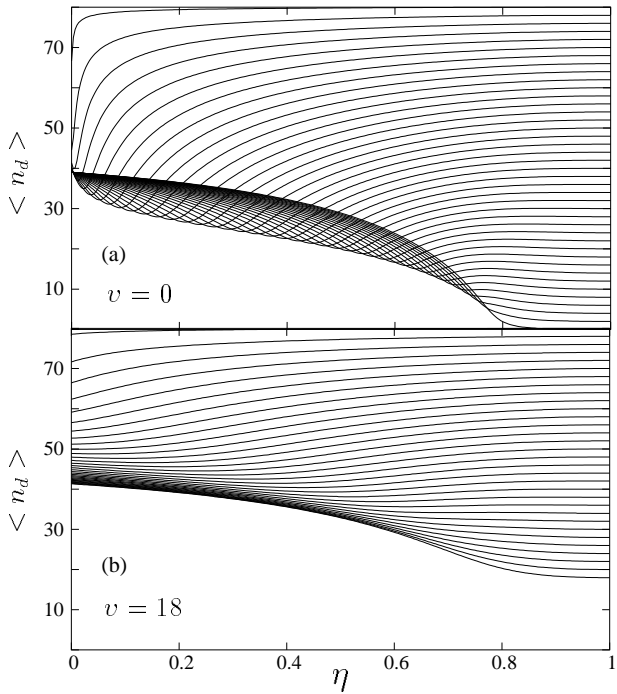,width=\linewidth}
\caption{\protect\small The average number of $d$-bosons for $v=0$ 
(panel a) and $v=18$ (panel b) states with $l=0$ and $N=80$.}
\label{ndav}
\end{figure}

From Eq.~(\ref{veloc}) we see that $(\eta_{\rm f},E_{\rm f})$ will be
a focal point of Hamiltonian (\ref{hamgen}) if 
$\matr{\psi_i(0)}{\hat{H}(\eta_{\rm f})}{\psi_i(0)}=E_{\rm f}$,
so in our particular case we have
\begin{equation}
\matr{\psi_i(0)}{\hat{n}_{d}}{\psi_i(0)}\approx\frac{N}{2}\ ,
\end{equation}
where $\ket{\psi_i(0)}$ are the Hamiltonian eigenvectors with $v=0$
at $\eta=0$.
This means that the average number of $d$-bosons in individual
O(6) eigenstates with zero seniority stays nearly constant across
the whole spectrum. 
Figure~\ref{ndav}(a), where the $n_d$ average is shown explicitly along 
the whole $\eta\in[0,1]$ path for all $v=0$ levels with $N=80$, supports 
this rule; see the $\eta=0$ limit (graphically it is difficult to 
distinguish, whether the convergence of all curves to the $\frac{N}{2}$ 
point is exact or not, but numerical values indicate that it is only 
approximate).
With a lower precision, the validity of the above
\lq\lq spectral invariant\rq\rq\ can be extended to higher seniorities,
but with increasing $v$ there are more and more upper states that do not 
fit, see Fig.~\ref{ndav}(b) that shows $\langle n_d\rangle_i$ for
the $v=18$ levels.

Even in the SU(3)$-$U(5) and ${\overline{\rm SU(3)}}-$U(5) transitions, 
when $\hat{Q}$ in Hamiltonian (\ref{ham}) is replaced by $\hat{Q}_{\chi}$ 
with $\chi=-\frac{\sqrt{7}}{2}$ or $\chi=+\frac{\sqrt{7}}{2}$ and the
seniority is not conserved, one finds a similar approximate invariant, 
namely $\matr{\psi_i(0)}{\hat{n}_{\rm d}}{\psi_i(0)}\approx\frac{3N}{4}$, 
where $\ket{\psi_i(0)}$ represent the SU(3) or ${\overline{\rm SU(3)}}$ 
eigenvectors (the complete $l=0$ spectrum for these transitions can be 
found in Ref.~\cite{Cej4}).

Note that we first detected these invariants geometrically, from the
focal points.
The impact of such invariants on the level dynamics can be enormous
since focal points represent an essential condition for the initial 
compression of the Coulomb gas, which results in stronger interactions
between levels.
This compression triggers the formation of the \lq\lq shock wave\rq\rq\ 
in the densest part of the spectrum, see Sec.~\ref{shock}.
Therefore, the existence of an initial (exact or approximate) focal 
point may belong to the main causes that eventually lead to a phase 
transition at some point.

\subsection{Finite-$N$ phase transitions}
\label{finN}

Although we are dealing here with the spherical-deformed transition
induced by varying parameter $\eta$ in Hamiltonian (\ref{ham}), one
should realize that the [O(6)$-$U(5)]$\supset$O(5) transitional path 
itself coincides with the separatrix between prolate and oblate 
deformed phases \cite{Mor,Jol1}.
The prolate-oblate first-order phase transition for $N\to\infty$ at 
any fixed value of $\eta\in[0,\frac{4}{5})$ can be induced by 
varying parameter $\chi$ in the generalized Hamiltonian of the form 
(\ref{ham}) with $\hat{Q}$ replaced by $\hat{Q}_{\chi}$.

It is argued in Ref.~\cite{Ari} that the O(6) dynamical symmetry 
represents a special point of the IBM phase diagram where 
a discontinuous prolate-oblate change of the ground-state structure 
can be observed even for {\em finite\/} boson numbers.
Indeed, if one explicitly includes the O(5) Casimir invariant 
(\ref{senior}) into the Hamiltonian with a coefficient such that 
the $v=0$ ground-state at $(\eta,\chi)=(0,0)$ becomes degenerate 
with the lowest states of other seniorities, a crossing of the 
ground-state configurations will occur for any value of $N$ when 
passing the O(6) point in the $\chi$-direction.

It is easy to show that this mechanism can be extended to the whole 
$\eta\in[0,\frac{4}{5})$ transitional region.
The basic trick---the fact that levels with different seniorities
can be made degenerate---remains the same.
After subtracting the component corresponding to the O(5) Casimir 
invariant \cite{Cej6} from the general $\chi$-dependent Hamiltonian 
of the form (\ref{ham}), one arrives at the expression
\begin{equation}
\hat{H'}(\eta,\chi)\propto\frac{\eta-1}{N^2}\left\{
(\hat{Q}_{\chi}\cdot\hat{Q}_{\chi})-\frac{1}{2}\hat{C}_2[O(5)]\right\}
+\frac{\eta}{N}\,\hat{n}_{d}
\ ,
\label{hamdeg}
\end{equation}
that exhibits the desired property:
For any fixed value of $\eta$ and any finite boson number $N$, the 
ground state as a function of $\chi$ changes discontinuously at
$\chi_{\rm c}=0$, which thus defines the critical point for
a finite-$N$ prolate-oblate phase transition.

We therefore extend the region of possible finite-$N$ phase 
transitions in the IBM phase diagram to the whole prolate-oblate 
separatrix.
The key feature needed for this generalization of Ref.~\cite{Ari} is 
the integrability of the [O(6)$-$U(5)]$\supset$O(5) Hamiltonians.
Note, however, that phase transitions at finite dimensions, induced 
by unavoided crossings of levels involving the ground-state, 
represent a rather nongeneric kind of behavior, which moreover 
is not robust enough to survive at finite temperatures.
Indeed, if the temperature increases from zero to an infinitesimally 
small value, nonzero populations of both levels result in a smooth
dependence of the free energy on the control parameter, and the 
phase-transitional behavior is washed out.

\subsection{Bulk properties of the spectrum}
\label{bulk}

It is clear that the strongest influence on a given level comes
typically from its neighbors at the places of avoided crossings.
Besides these binary interactions (involved in the shock-wave 
propagation), there exists also a component of 
the total force acting on each individual level that originates from 
the bulk of the whole ensemble.
In this subsection, we will consider two global measures of this
bulk component.

First, we consider the overall compression of all levels, represented 
by the energy dispersion (squared \lq\lq spread\rq\rq) of the spectrum, 
$\Delta_E^2=\frac{1}{n}\sum_{i}(E_i-{\bar E})^2$, where 
${\bar E}=\frac{1}{n}\sum_{i} E_i$ is a center-of-mass energy. 
A straightforward calculation yields the expression
\begin{eqnarray}
\Delta_E^2=
\biggl[\frac{{\rm Tr}\hat{H}_0^2}{n}-\frac{{\rm Tr}^2\hat{H}_0}{n^2}\biggr] 
+
2\eta\biggl[\frac{{\rm Tr}(\hat{H}_0\hat{V})}{n}-\frac{{\rm Tr}\hat{H}_0
{\rm Tr}\hat{V}}{n^2}\biggr]
\nonumber\\
+ \eta^2\biggl[\frac{{\rm Tr}\hat{V}^2}{n}-\frac{{\rm Tr}^2\hat{V}}
{n^2}\biggr]
\ ,
\qquad\ 
\label{parab}
\end{eqnarray}
which shows that the spectral dispersion is a quadratic function with a 
minimum at
\begin{equation}
\eta_0=-\frac{n{\rm Tr}(\hat{H}_0\hat{V})-{\rm Tr}\hat{H}_0{\rm Tr}\hat{V}}
{n{\rm Tr}\hat{V}^2-{\rm Tr}^2\hat{V}}
\label{mini}
\ .
\end{equation}
For $\eta\approx\eta_0$, the strengths of both terms $\hat{H}_0$ and 
$\eta\hat{V}$ of Hamiltonian (\ref{hamgen}) are comparable, so that  
the strongest effects of mixing take place in the surrounding region.
For $\eta\gg\eta_0$ or $\eta\ll\eta_0$, on the other hand, the spectrum 
just blows up, the Hamiltonian being dominated by $\eta V$.

\begin{figure}
\epsfig{file=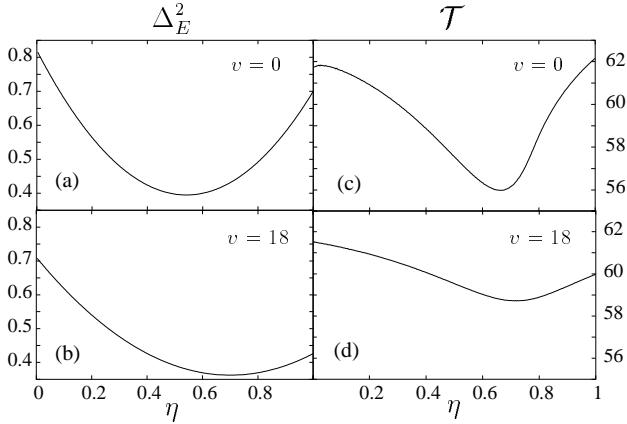,width=\linewidth}
\caption{\protect\small The dispersion (\ref{parab}) of the spectrum and 
the kinetic energy from Eq.~(\ref{toten}) for $v=0$ and 18 states,
corresponding to $l=0$ and $N=80$.}
\label{bulky}
\end{figure}

For (a) $v=0$ and (b) $v=18$ subsets of the spectrum with $N=80$, 
the function (\ref{parab}) is shown in the two leftmost panels of 
Figure~\ref{bulky}.
We see that the $v=0$ levels are maximally compressed at $\eta_0\approx 
0.56$, i.e., in the region just before the phase transition.
For higher seniorities, the minimum moves towards $\eta_{\rm c}=
\frac{4}{5}$.
Let us note that a similar conclusion can be made for $\chi\neq 0$, when
of course the seniority is not conserved and the contribution of all 
$l=0$ levels must be summed up.
For $\chi=\pm\frac{\sqrt{7}}{2}$, for instance, the energy dispersion 
forms a sharp minimum directly at $\eta_0\approx 0.8$.

The second quantity we will use here to characterize the bulk component
of the force is the total product charge ${\cal Q}=\sum_{i>j}Q_{ij}=
\sum_{i>j}|V_{ij}|^2$. 
It is related to the sum
\begin{equation}
\underbrace{\frac{1}{2}\sum_i \left( \frac{dE_i}{d\eta} \right)^2}_{\cal T}
+\underbrace{\frac{1}{2}\sum_{i\neq j}|V_{ij}|^2}_{{\cal V}}=\frac{1}{2}
{\rm Tr}\hat{V}^2\equiv{\cal E}
\label{toten}
\ ,
\end{equation}
which is an integral of motions of the Pechukas-Yukawa model, known as the 
total energy \cite{Stock}.
Since $\frac{d}{d\eta}{\cal E}=0$, the second term that represents the
potential energy ${\cal V}={\cal Q}$ is at any value of $\eta$ just 
a complement of the first, kinetic term ${\cal T}$.
For $\eta\gg\eta_0$ and $\eta\ll\eta_0$ [assuming for a while 
$\eta\in(-\infty,+\infty)$], the eigenbasis of $\hat{H}(\eta)$ virtually 
coincides with the eigenbasis of $\hat{V}$ so that 
${\rm Tr}\hat{V}^2\approx\sum V_{ii}^2=2{\cal T}$ and ${\cal V}\approx 0$.
In these regions, the gas just freely expands.   
On the other hand, around $\eta_0$ the kinetic and potential terms in 
Eq.~(\ref{toten}) are comparable and the interaction may generate 
nontrivial effects.

The kinetic energy from Eq.~(\ref{toten}) for levels with $v=0$ 
and $v=18$, respectively, is shown in panels (c) and (d) of 
Fig.~\ref{bulky}.
In both cases, we observe a minimum of $\cal T$ very close to the 
critical point; for $v=0$ the minimum is located at $\eta\approx 0.67$.
This means that ${\cal V}={\cal Q}$ is maximal at the same place,
implying the strongest overall strength of level interactions.
For higher seniorities, the minimum gets shallower and moves
towards $\eta_{\rm c}$.

We saw that both the compression of the spectrum and total interaction 
strength are maximal in the region of control parameters around 
$\eta_0$ which immediately precedes the phase transition at 
$\eta_{\rm c}$.
Conversely, Eq.~(\ref{mini}) yields a reasonable rough estimate of the 
parameter range of a general Hamiltonian (\ref{hamgen}) where eventual 
phase transitions may be located.

\section{Eigenstate dynamics}
\label{state}

Besides dynamics of individual Hamiltonian eigenvalues $E_i(\eta)$, one 
can also analyze structural changes of the corresponding eigenstates 
$\ket{\psi_i(\eta)}$.
These two aspects of spectral evolution are mutually correlated, since 
the matrix elements $V_{ij}$, that carry information on wave 
functions, belong to dynamical variables involved in Pechukas-Yukawa 
equations (\ref{force})--(\ref{veloc}).

\begin{figure}
\epsfig{file=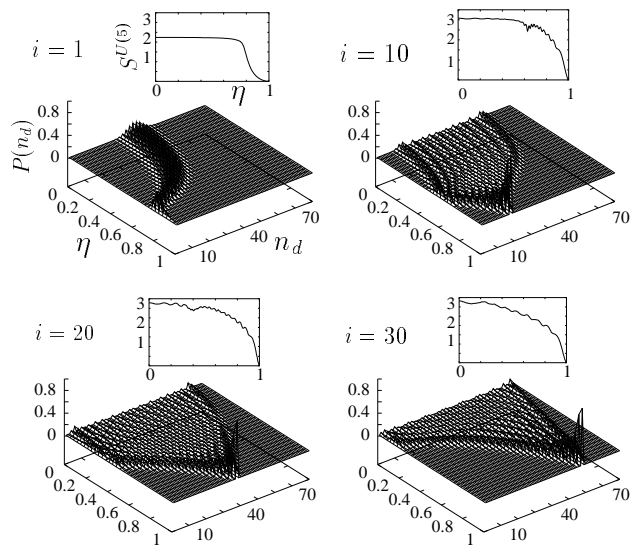,width=\linewidth}
\caption{\protect\small The distribution of the $d$-boson number $n_d$ 
 in four $l=v=0$ eigenstates of Hamiltonian (\ref{ham}) with $N=80$ as 
 a function of $\eta$. The insets show the corresponding U(5) 
 wave-function entropy.}
\label{wavef}
\end{figure}

In Fig.~\ref{ndav}, we have already seen the evolution of the average
number of $d$-bosons, $\langle n_d\rangle_i$, in the $v=0$ and $v=18$ 
eigenstates.
This information is now supplemented by Figure~\ref{wavef}, where the 
$\eta$-dependence of the whole distribution $P_i(n_d)$ of $n_d$ is 
shown for selected $N=80$ Hamiltonian eigenstates, namely the $v=0$ 
states with $i=1$, 10, 20, and 30 (ordered with increasing energy). 
For $l$ and $v$ fixed, the probability
\begin{equation}
P_i(n_d)\bigr|_{\eta}=\sum_{{\tilde n}_{\Delta},m}
|\scal{n_d,v,{\tilde n}_{\Delta},l,m}{\psi_i(\eta)}|^2
\end{equation}
for each $\eta$ is determined as the projection of the state 
$\ket{\psi_i(\eta)}$ onto the subspace of the U(5) eigenstates 
with $n_d$ equal to the given number.
In the U(5) limit, the distribution is concentrated on a single value 
$n_d=2i-2$ (with zero seniority, the value of $n_d$ must be even), 
but it quickly spreads over a broad range of $n_d$'s as $\eta$ 
decreases from 1 to 0.

Figure~\ref{wavef} shows four qualitatively different types of ways
how this delocalization proceeds:
For the ground state, $i=1$, the value of $n_d$ remains zero as far 
as $\eta>\eta_{\rm c}$, and then it suddenly increases (with decreasing
$\eta$), forming a ridge around the average that goes approximately
as $\langle n_d\rangle_1\propto\sqrt{\eta_{\rm c}-\eta}$, in agreement 
with the phase-transitional predictions; cf. Fig.~\ref{ndav}(a).
For excited states, the gradual spread of wave functions in $n_d$ can 
be compared to the propagation of waves on a string.
The string is initially (at $\eta=1$; the \lq\lq time\rq\rq\ is now 
thought to go backwards) subject to an instantenous point perturbation 
and the resulting waves propagate in both $n_d=0$ and $n_d=N$ 
directions asymmetrically.
The pattern of wave propagations changes with $i$: for instance, the 
speed of the upper wave is lower for higher excited states.
When the lower front of the wave reaches the $n_d=0$ limit, it either
gets reflected (this happens for lower excited states, see the $i=10$
example) or stops there (for higher excited states, see $i=20$ and 30
cases).

It is interesting that the value of $\eta$ where the wave reaches the 
lower endpoint $n_d=0$ coincides with the range where the \lq\lq shock 
wave\rq\rq\ affects the given level, see Figs.~\ref{levsen}(a) and
\ref{distan}(a).
This can be checked for a larger set of levels in Fig.~\ref{ndav}(a). 
The dependences of individual $n_d$ averages exhibit well-pronounced 
minima that correspond to the stopping or reflection of the lower
wave front at $n_d=0$, and reasonably coincide with the moments of 
passage of the shock-wave.

Also shown in the insets of Fig.~\ref{wavef} is the U(5) wave-function 
entropy, 
\begin{equation}
S_i^{\rm U(5)}=-\sum_{n_d=0}^{N}P_i(n_d)\ln P_i(n_d)\ ,
\end{equation}
which measures the overall spread of the instantenous eigenvector 
$\ket{\psi_i(\eta)}$ in the U(5) basis \cite{Cej6}.
Assuming a quasiuniform distribution of the $i$th state over a certain
set of the $\hat{n}_d$ eigenstates, one finds that the effective number
of components is given by $n_i^{\rm eff}=\exp S_i^{\rm U(5)}$.
This number is approximately equal to half of the width (at a given 
value of $\eta$) of the $n_d$ distribution corresponding to the 
respective level (taking into account that odd $n_d$ values are not 
populated for $v=0$).

As can be seen in Fig.~\ref{wavef}, the widths of the $n_d$ distributions
and the corresponding U(5) entropies grow with decreasing $\eta$ as far as 
the distribution touches the $n_d=0$ limit (the level gets into the 
shock-wave region). 
After this point, the width and entropy stay approximately constant.
If proceeding from the O(6) side, i.e., returning to the forward direction
of \lq\lq time\rq\rq, we can say that the process of localization of 
level $i$ in the U(5) basis {\em starts\/} approximately when the shock 
wave hits the level.
This supports and further specifies the shock-wave scenario described
in Sec.~\ref{shock}.
We must stress, however, that for excited states the transition to the 
U(5) structure after passing the shock wave is only gradual. 
A sudden phase-transitional type of change is reserved for the ground 
state only.

\begin{figure}
\epsfig{file=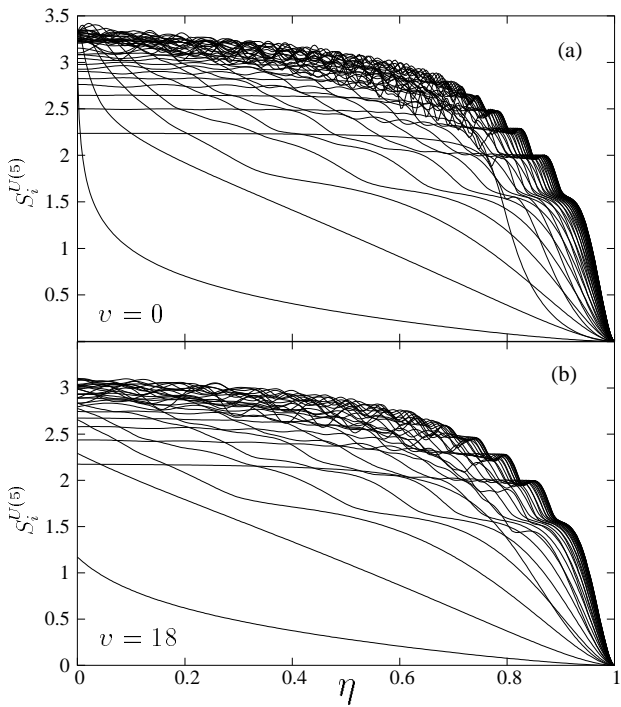,width=\linewidth}
\caption{\protect\small The U(5) wave function entropy for all 
 $v=0$ (panel a) and $v=18$ (panel b) eigenstates of Hamiltonian 
 (\ref{ham}) with $l=0$ and $N=80$.}
\label{entrop}
\end{figure}

As indicated by the $i>1$ examples in Fig.~\ref{wavef}, the decrease 
of the U(5) wave-function entropy exhibits some undulations, connected 
with quantum interferences of the amplitudes corresponding to 
populations of individual $n_d$'s.
It is surprising that vertical coordinates of the main oscillations 
are about constant for the whole ensemble of states.
This is demonstrated in Figure~\ref{entrop}, where we show the U(5) 
wave-function entropy for all (a) $v=0$ and (b) $v=18$ states ($N=80$).
Clearly, if one proceeds from state to state, the undulations are 
shifted in $\eta$, but remain at about the same levels of entropy.
The result is a peculiar pattern of \lq\lq plateaus\rq\rq\ present 
in both panels of Fig.~\ref{entrop}. 
(Let us stress, however, that these plateaus are only a visual 
effect appearing when all entropies are drawn in the same figure.)
This hints at strong correlations in the structural changes of 
individual eigenstates after the passage through the shock-wave
region.

The most distinguished steps of the patterns in Fig.~\ref{entrop} 
are the same for both seniorities. 
They correspond to the effective numbers of wave-function components 
equal approximately to $n^{\rm eff}_i\approx 4.5$, 7.5, 10, and 12.
Note that the average delocalization of a given state in a randomly 
chosen basis is for sufficiently high dimensions $n$ given by 
$n^{\rm eff}_{\rm GOE}\approx 0.48\,n$ \cite{Cej6}, which for the $v=0$ 
and $v=18$ subspaces yields typical saturation values of the wave-function
entropy equal to $S_{\rm GOE}\approx 3$ and $\approx 2.7$, respectively.
(The largest U(5) entropies in the $\eta=0$ limit slightly exceed the
GOE values, but the latter provide reasonable estimates of averages 
if all states are taken into account.)
We see that the system of plateaus in Fig.~\ref{entrop} disappear in 
noisy oscillations just below the respective GOE entropy values.

Let us stress that no step-like structures are observed in cumulative 
plots of the U(5) wave-function entropy of all $l=0$ states for the 
SU(3)$-$U(5) and ${\overline{\rm SU(3)}}-$U(5) transitions.
The present correlated behavior is therefore connected solely with 
the integrable $\chi=0$ region.

\section{Conclusions}
\label{conclu}

We have studied dynamics of the $l=0$ energy levels and the 
corresponding eigenstates along the [O(6)$-$U(5)]$\supset$O(5) 
transition of the interacting boson model.
Results of our numerical calculations were discussed in the 
framework of the Pechukas-Yukawa model, which describes the evolution 
of quantal spectra as one-dimensional motions of an ensemble of 
classical particles.
Treated in this way, spectral attributes for all values of the 
control parameter---including possible phase transitions at some 
critical points---result just from a specific \lq\lq initial 
condition\rq\rq, i.e., the set of energies and interaction matrix 
elements at a single {\em arbitrary\/} point $\eta$.
Of course, particularly tempting is to consider the whole spectral
evolution along $\eta\in[0,1]$ (and beyond) being predetermined by 
properties of the system in either of the two limiting dynamical 
symmetries.

We disclosed cooperative and highly coherent behaviors of the 
individual spectral constituents, i.e., level energies and wave 
functions corresponding to various seniorities.
This may be generally linked to the integrability of the model 
in the present regime, namely to the possibility to separate
seniorities, but we have to admit that some of the findings 
remain just plain observations.
Further studies may still shed more light on how this all 
\lq\lq comes about\rq\rq.

The most significant cooperative effect seems to rely on the
\lq\lq shock-wave\rq\rq\ mechanism, that consists in an ordered 
sequence of avoided crossing of levels in the region around 
$E\approx 0$ and the accompanying changes of eigenstates 
(Secs.~\ref{shock} and \ref{state}).
Triggered by an initial compression of the spectrum, the shock 
wave initiates in its densest upper part and propagates 
downwards to the ground state.
The passage of the wave through a given state starts the gradual
transfiguration of the state structure into the U(5) form.
This mechanism provides a deeper insight into the process that 
eventually leads to the ground-state phase transition of second 
order.

Among the other findings we highlight the following: (a) approximate 
focal points of IBM spectra in transitions to the U(5) dynamical 
symmetry (Sec.~\ref{focal}), (b) possible finite-$N$ prolate-oblate 
phase transitions along the whole [O(6)$-$U(5)]$\supset$O(5) 
separatrix (Sec.~\ref{finN}), (c) extremes of spectral \lq\lq bulk 
observables\rq\rq\ in the region immediately preceding the phase 
transition (Sec.~\ref{bulk}), (d) highly correlated changes of 
consecutive eigenstates leading to \lq\lq plateaus\rq\rq\ in the 
cumulative plot of U(5) wave-function entropies (Sec.~\ref{state}).

In the following part of our article \cite{partII}, we will
focus on the interpretation of the $E\approx 0$ pattern of
level bunchings (Sec.~\ref{bunch}) within the semiclassical
theory of quantal spectra.

\acknowledgments

We acknowledge useful discussions with R.F.~Casten and J.~Dobe{\v s}.
This work was supported by the DFG grant no.\,U36 TSE 17.2.04.

\end{document}